\documentclass[twocolumn,english,superscriptaddress,aps]{revtex4}
\usepackage[latin9]{inputenc}
\usepackage{amsmath}
\usepackage{amssymb}
\usepackage{graphicx}
\usepackage{color}

\makeatletter

\@ifundefined{textcolor}{}
{%
 \definecolor{BLACK}{gray}{0}
 \definecolor{WHITE}{gray}{1}
 \definecolor{RED}{rgb}{1,0,0}
 \definecolor{GREEN}{rgb}{0,1,0}
 \definecolor{BLUE}{rgb}{0,0,1}
 \definecolor{CYAN}{cmyk}{1,0,0,0}
 \definecolor{MAGENTA}{cmyk}{0,1,0,0}
 \definecolor{YELLOW}{cmyk}{0,0,1,0}
}

\newcommand{\ket}[1]{\left|#1\right>}
\newcommand{\bra}[1]{\left<#1\right|}
\newcommand{\ii}{\mathrm{i}}
\newcommand{\f}[1]{\mbox{\boldmath$#1$}}
\newcommand{\nn}{\nonumber\\}
\newcommand{\trace}[1]{{\rm Tr}\left\{#1\right\}}

\newcommand{\abs}[1]{\left| {#1} \right|}

\makeatother

\usepackage{babel}
\begin{document}

\title{Cooperative efficiency boost for quantum heat engines}

\author{David Gelbwaser-Klimovsky}

\affiliation{Department of Chemistry and Chemical Biology, Harvard University,
Cambridge, MA 02138}

\author{Wassilij Kopylov}

\affiliation{Institut fur Theoretische Physik, Technische Universit\"at Berlin, D-10623 Berlin, Germany}

\author{Gernot Schaller}

\affiliation{Institut fur Theoretische Physik, Technische Universit\"at Berlin, D-10623 Berlin, Germany}

\begin{abstract}
The power and efficiency of many-body heat engines can be boosted by performing cooperative non-adiabatic operations in contrast to the commonly used adiabatic implementations. 
Here, the key property relies on the fact that non-adiabaticity is required in order to allow for cooperative effects, that can use the thermodynamic resources only present in the collective non-passive state of a many-body system. 
In particular, we consider the efficiency of an Otto cycle, which increases with the number of copies used and reaches a many-body bound, which we discuss analytically.  
\end{abstract}
\maketitle

{\bf Introduction} -- From a classical perspective, adiabatic processes are those that do not inject heat into the system~\cite{landau1980statistical}. 
Within the context of heat engines~\cite{kosloff2013quantum,gelbwaser2015thermodynamics,allahverdyan2008work} 
this means that by maximizing the energetic variation of an adiabatic driving protocol, 
the work output is optimized. 
Adiabatic processes are typically associated with slow transformations, where slow is defined in term of the system equilibration time, such that
an adiabatic process could be actually quite fast, as long as the system is thermally isolated. 
At the quantum level, the question of the adiabatic process speed is answered by the quantum adiabatic theorem (QAT), e.g.~\cite{berry1984a,wu1989a,sarandy2004a,schaller2006b,jansen2007a}. 
The QAT warrants that a (typically slow enough) Hamiltonian transformation results in a quantum adiabatic (QA) process, defined by constant populations of time-dependent energy levels.
However, for any unitary (and thereby isentropic) transformation, any energy difference resulting from it can be reversed and therefore does not involve irreversible losses.

In this letter, we discuss an equality for the efficiency with quantum-information measures which shows that  cooperative many-body heat engines that violate the QAT by a non-QA (NQA) process during their isentropic strokes can have larger efficiency and power than their slow QA or non cooperative counterparts~\cite{zheng2014work,quan2007quantum}. 
In contrast to 
other  methods to increase the power or the 
efficiency~\cite{jaramillo2016quantum,hardal2018phase,niedenzu2018cooperative,
niedenzu2015performance,del2014more,halpern2017mbl}, our results 
are based on the non-passivity of the collective state for which QA transformations are suboptimal for work extraction and efficiency.

\begin{figure}
	\centering
	\includegraphics[width= \columnwidth ]{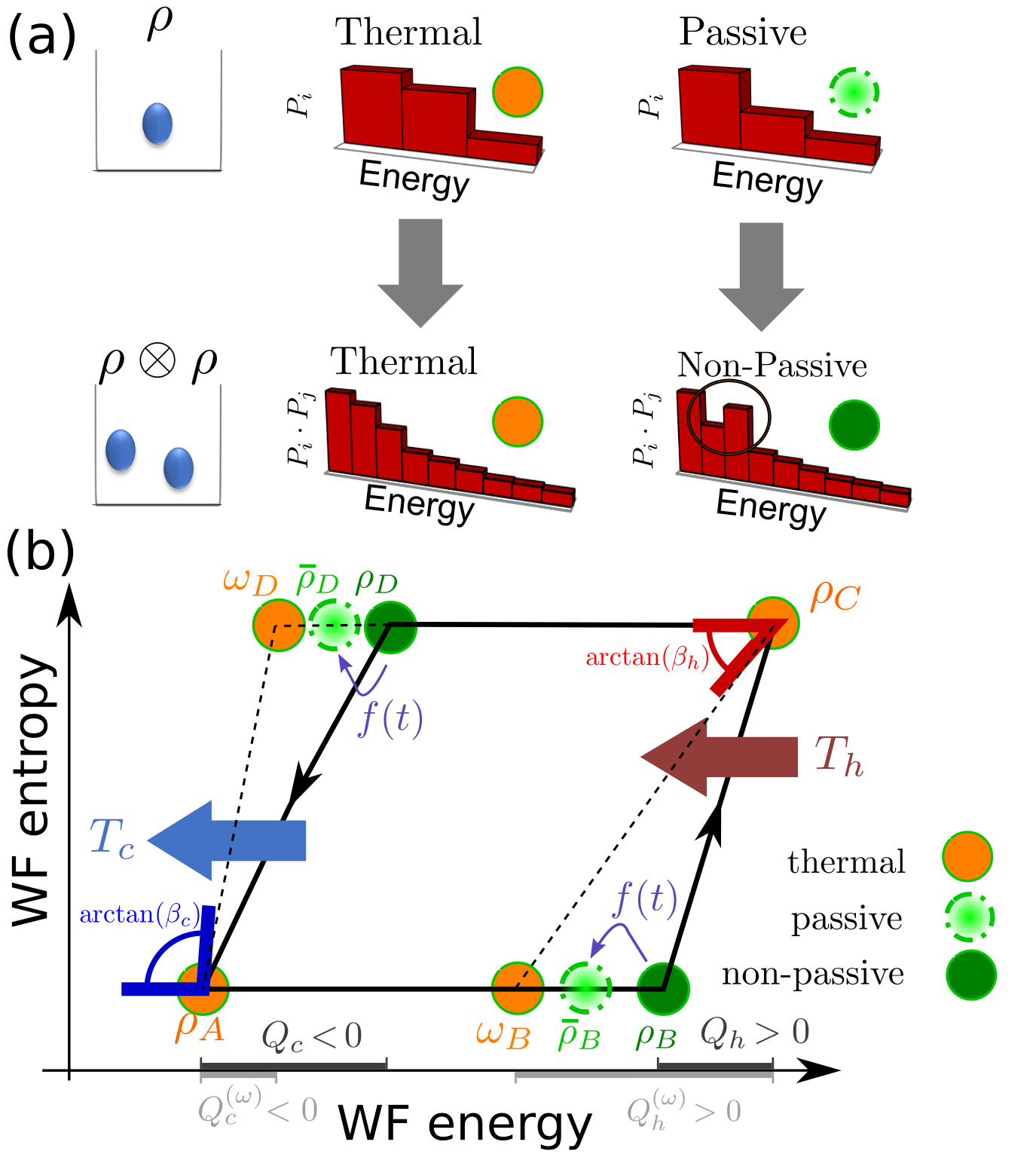}
	\caption{(a) Multiple copies (left) of a thermal state always create a collective thermal state (middle), whereas multiple copies of a passive state may produce a collective non-passive state (right). 
	(b) A non-interacting many body system undergoes an Otto cycle (black closed loop) achieving suboptimal work extraction and efficiency due to non-passive states (green solid circles). When interactions, denoted as $f(t)$, are allowed, passive states (dash-dotted cirlces) are created at $B$ and $D$, reducing the energy at these points and increasing the work extraction by changing the heat exchange (bottom axis) and efficiency. The thermal reference states $\omega_B$ and $\omega_D$ (orange solid circles) define a working cycle with optimal efficiency at fixed Hamiltonians at $B$ and $D$ (dashed lines, see Appendix C), providing a tighter bound than the second law
(angles ($\sphericalangle$) at $A$ and $C$, see Appendix B). 
}
	\label{fig:fig1}
\end{figure}

The analyzed effect can be understood from two basic properties of QA processes performed on an initially thermal system: i) for a single system an inhomogeneous shift of the energy levels~\cite{gelbwaser2018single,levy2018quantum} produces a non-thermal state with zero ergotropy~\cite{allahverdyan2004maximal,niedenzu2016operation}, i.e., for cyclic Hamiltonian modulations, its energy cannot be reduced by a unitary transformation. 
Thus, work cannot be extracted from it unless they are coupled to a non-equilibrium system (e.g. two thermal baths at different temperatures or chemical potentials) or the Hamiltonian is modified in a non-cyclic way. 
This state is called {\em passive}~\cite{pusz1978passive,lenard1978thermodynamical,gelbwaser2013work,allahverdyan2004maximal}; 
ii) A collective state formed by 
identical copies of a passive state  may be non-passive~\cite{pusz1978passive} (see Fig.~\ref{fig:fig1}(a)), i.e., it may have positive ergotropy and therefore extractable work.

In order to exploit the extra work that resides in the collective nature of the system, the Hamiltonian modulation should be a NQA process, allowing population transfers among different copies of the working fluid (WF), such that the collective state at the end of the isentropic stroke is passive. 
This strategy reduces the stroke time and increases the extracted work, thereby boosting both power and  efficiency, which is still limited by the Carnot bound (see Appendixes A and B).

{\bf Passive states}--  
We start by analyzing an isolated quantum  system initially at thermal equilibrium that undergoes a 
QA Hamiltonian transformation. 
Assuming that the system has discrete energy levels, the  initial population of level $n$ is $P_n(0)=Z^{-1} e^{-\beta E_n(0)}$, where $E_n(0)$ is the initial energy of level $n$, $\beta=\frac{1}{k_B T}$ and $Z$ is a normalization constant.
If the energy levels are altered homogeneously (i.e., for every level $E_n(t_f)=q E_n(0)$, where $q > 0$ is the same for all the levels and the process ends at time $t_f$), the final QA state would be a thermal state of the final energies, $E_n(t_f)$, but with inverse temperature $\frac{\beta}{q}$, i.e., $P_n(t_f)=Z^{-1} e^{-\frac{\beta}{q} E_n(t_f)}$. 
For a general modulation, the energy level scaling is not homogeneous. 
As it has been shown in~\cite{gelbwaser2018single}, for inhomogeneous energy level transformations classically inconceivable heat engines can be realized such as those operating with an incompressible working fluid. 
Under this type of transformations, the final state is not thermal but passive. 
In practice, passivity means that the state is diagonal in the eigenenergy basis and the populations of the levels decrease with the energy without necessarily following a Boltzmann distribution. 

Quantum and classical systems can both be at passive states~\cite{gorecki1980passive}, which we denote
by an overbar (e.g. $\bar\rho$).
Nevertheless, their behavior drastically diverges when we consider several copies of the same passive state.  
For {\it classical} systems -- having a continuous energy spectrum -- a passive state $\bar{\rho}^{\rm cl}$, 
whose two-copies-state $\bar{\rho}^{\rm cl}_2=\bar{\rho}^{\rm cl}\otimes\bar{\rho}^{\rm cl}$ is also  passive, 
always produces a multi-copy passive state $\bar{\rho}^{\rm cl}_N=\otimes_N \bar{\rho}^{\rm cl}$. 
In contrast,  {\it quantum} systems -- having discrete energy levels -- are different: 
$\otimes_N \bar{\rho}^{\rm qm}$ could be non-passive independently of the passivity for $N=1,2$~\cite{gorecki1980passive}.

{\bf Quantum Otto cycle.} -- We study the Otto cycle~\cite{zheng2014work,quan2007quantum,rossnagel2016single,newman2017a}, 
see Fig.~\ref{fig:fig1}(b), undergone by a WF. 
It has 4 stages: 
At point A its Hamiltonian is $H_A$ and its $n$th energy level is $E_{A,n}$. 
The WF is at thermal equilibrium with the cold bath at inverse temperature $\beta_c$, i.e., 
$\rho_{A}\propto e^{-\beta_c H_{A}}$. 
The {\it isentropic ``compression'' } stroke connects $A$ and $B$: The WF is isolated and the Hamiltonian becomes time-dependent. 
It generates the Hamiltonian $H_{B}$, with tracked energies denoted by $E_{B,n}$. 
Since we also consider NQA protocols, in general $\rho_B$ could have different populations than $\rho_A$
in the eigenbases of $H_B$ and $H_A$, respectively.
The second stroke, {\it hot equilibration},  involves a constant Hamiltonian and the coupling of the WF to a hot bath, with inverse temperature $\beta_h$, until the WF reaches thermal equilibrium at the bath temperature. 
This point is denoted as C and $\rho_C\propto  e^{-\beta_h H_{C}}$ with $H_{C} = H_B$. 
The third stroke is the {\it isentropic ``expansion''}, where the WF is isolated again, and the Hamiltonian $H_{B}$ is transformed to $H_{D} = H_{A}$ in a reversed fashion. 
This stroke ends at point D, where the populations of $\rho_{D}$ in general differ from  $\rho_C$. 
The {\it cold equilibration} stroke closes the cycle by coupling the WF to the cold thermal bath until it thermally equilibrates with it, returning to point A.

The work per cycle is the exchanged energy during the isentropic strokes, i.e.,
\begin{equation}
\label{eq:work}
W= {\rm Tr}\lbrack H_{B} \rho_B\rbrack -{\rm Tr}\lbrack H_{A} \rho_{A}\rbrack+  
{\rm Tr}\lbrack H_{A} \rho_D\rbrack - {\rm Tr}\lbrack H_{B} \rho_{C}\rbrack,
\end{equation}
and the heats exchanged with the hot and cold bath are 
\begin{align}
\label{eq:heat_eq}
Q_h= {\rm Tr}\lbrack H_B (\rho_C-\rho_B)\rbrack;  \
Q_c= {\rm Tr}\lbrack H_A (\rho_A-\rho_D)\rbrack.
\end{align}
The sign convention is that energy flowing in (out) of the WF is positive (negative). 
The efficiency is the ratio between work extracted  $(W < 0)$ and incoming heat, i.e., $\eta=-W/Q_h=1- |Q_c|/Q_h$. 
Thus, a smaller heats ratio $|Q_c|/Q_h$, results on a larger efficiency, which is bounded by Carnot efficiency $\eta_C$.   

Performing QA transformations that homogeneously scale the energy levels by construction produces thermal states at the end of the isentropic strokes, which correspond to the minimal energy state of an isentropic process. 
In general, minimizing the energy at points B and D of the cycle  optimizes the extracted work and efficiency.  
But if instead the energy levels are inhomogeneously scaled, the state of a many body WF after a QA transformation could diverge from the minimal energy state at constant entropy.

As shown in the Appendix C, the full efficiency of the Otto cycle is given by 
\begin{equation}
\label{eq:full_efficiency}
\eta = 1 - \left[1 + \frac{\mathcal{D}(\rho_D||\omega_D)}{\beta_D^{(\omega)} (-Q_{c}^{(\omega)})} \right] 
	\sum_{n=0}^{\infty} \frac{\mathcal{D}^n(\rho_B||\omega_B)}{\left[\beta_B^{(\omega)} Q_{h}^{(\omega)}\right]^n}
	\left[\frac{-Q_{c}^{(\omega)}}{Q_{h}^{(\omega)}}\right]\,,
\end{equation}
where $\mathcal{D}(\rho||\omega) \geq 0$ is the quantum relative entropy between two density matrices.
The density matrix $\omega_{B(D)}$ denotes a constructed thermal density matrix with a 
{\it reference temperature} $\beta_{B(D)}^{(\omega)}$, uniquely defined for all density matrices 
with the same entropy $S$ and a Hamiltonian, via the Gibbs-state with the same von-Neumann entropy $S(\rho_{B(D)})=S(\omega_{B(D)})$~\cite{alicki2013entanglement}.
Consistently, $Q_{c(h)}^{(\omega)}$ is the corresponding cold (hot) heat using the constructed thermal density matrix $\omega_{B(D)}$ instead of $\rho_{B(D)}$ in~(\ref{eq:heat_eq}).
The values $Q_{c(h)}^{(\omega)}$do not depend explicitly on all properties of $\rho_D$ and $\rho_B$, but only on their entropies  as well as on the Hamiltonians at B and D.
As the von-Neumann entropy is constant under unitary operations $S(\rho_{B/D})=S(\rho_{A/C})$, the thermal reference heats and temperatures are the same for all protocols that connect the fixed Hamiltonians $H_{\nu}$, which allows to increase efficiency by minimizing the distances $\mathcal{D}(\rho_{B/D}||\omega_{B/D})$.
In the heat-engine regime one has $Q_{c}^{(\omega)} < 0 $  and $Q_{h}^{(\omega)} > 0$, 
and we see that approaching the reference states optimizes the heat exchanges, 
see bottom axis in Fig.~\ref{fig:fig1}(b). 
For a WF composed of a single copy, a QA protocol brings the WF to the respective passive state, 
$\bar{\rho}_{B(D)}$ which is the closest to $\omega_{B(D)}$ one can reach using unitary transformations.
For WF composed of multiple identical copies the collective energy levels may cross during the isentropic transformations.
Then, the QA protocol results in a  collective non-passive state $\rho_{B(D)}$. 
In comparison to a passive state $\bar{\rho}_{B(D)}$, such a non-passive state increases the quantum relative entropy, i.e., $\mathcal{D}(\rho_{B(D)}||\omega_{B(D)}) > \mathcal{D}(\bar{\rho}_{ B(D)}||\omega_{B(D)})$. 
As shown below, $\mathcal{D}$ can be decreased by performing swap operations which create a passive state. 

{\bf Single copy protocol} -- To illustrate different aspects of Eq.~\eqref{eq:full_efficiency}, we consider first  a simple example: a WF composed of a single three level system (qutrit, QT) with constant energy eigenstates $\{\ket{0},\ket{1}, \ket{2}\}$ and Hamiltonian
\begin{equation}\label{eq:single_copy}
H_{\rm QT}(t) = E_0 \ket{0}\bra{0} + E_1(t) \ket{1}\bra{1} + E_2 \ket{2}\bra{2}\,,
\end{equation}
where only the energy $E_1(t)$ is changed during the isentropic stages such that $E_0 < E_1(t) < E_2$.
Thus, even though the energy levels of a single QT do not cross, the modulation corresponds to an inhomogeneous scaling of the energy levels.
Further, since $[H_{\rm QT}(t),H_{\rm QT}(t')]=0$, it is always a QA transformation, and the final states at points $B$ and $D$ are not thermal, but passive and efficiency is maximized.
   
{\bf Multiple copy protocol} --  Now we consider that the WF is composed of several copies of the same system.   
Even though during the Hamiltonian modulation the energy levels of the single copy of the WF do not cross each other, the collective energy levels may do so.  
If this is the case, QA transformations (no transitions between smoothly connected eigenvalues) result in a collective non-passive state at points $B$ and $D$, even though the single system state is passive. 
%

As a concrete example, consider that -- despite non-crossing single system energy levels -- during the isentropic stroke two collective energy levels of the WF, $E_{\vec{n}=\{n_1,n_2,..\}}$ and $E_{\vec{m}=\{m_1,m_2,..\}}$, cross each other, where $n_i$ and $m_j \in \{0,1,2\}$. 
Particularly, assume that the collective Hamiltonian is 
\begin{gather}
\label{eq:Ham_two_QT}
H_N(t) = \sum_{i=1}^N H_{\rm QT}^i(t) +f(t) \left(\ket{\vec{n}}\bra{\vec{m}} + \ket{\vec{m}}\bra{\vec{n}}\right)\,,
\end{gather} 
where $H_{\rm QT}^i(t)$ is the $i$th copy single system Hamiltonian~(\ref{eq:single_copy}),
$f(t)$ is an additional interaction which we explain later in details  and $H_N(t_{\nu})=H_{\nu}$ for $\nu \in \{A,B,C,D\}$. 
At times when $f(t) = 0$, the collective energies are combinations of the energies of the single copy, i.e., $E_{\vec{n}=\{n_1,n_2,...\}} = \sum_i E_{n_i}$. 

For our example, we decompose the isentropic compression stroke into two substrokes.
In the first, the first excited state energy of the single system evolves for $t \in [t_A,t_{AB}]$ 
(with $t_A < t_{AB}<t_B$) as a linear ramp 
\begin{equation} 
\label{eq:E1_shift}
E_1(t) = E_1(t_A) + \Delta E_1 (t-t_{A})/(t_{AB}-t_A),  
\end{equation} 
and remains at $E_1(t_{AB}) = E_1(t_A)+\Delta E_1 = E_1(t_{B})$ (independent of $t_{AB}$) for the rest of the
full isentropic stroke.
Since this substroke is always of QA type, to speed up the protocol, $t_{AB} \to t_A$ can be assumed. 
The reverse transformation is performed during the isentropic expansion. 
In presence of collective level crossings, the final state of the isentropic substroke is non-passive.

\begin{figure}
	\centering
	\includegraphics[width=0.5\textwidth]{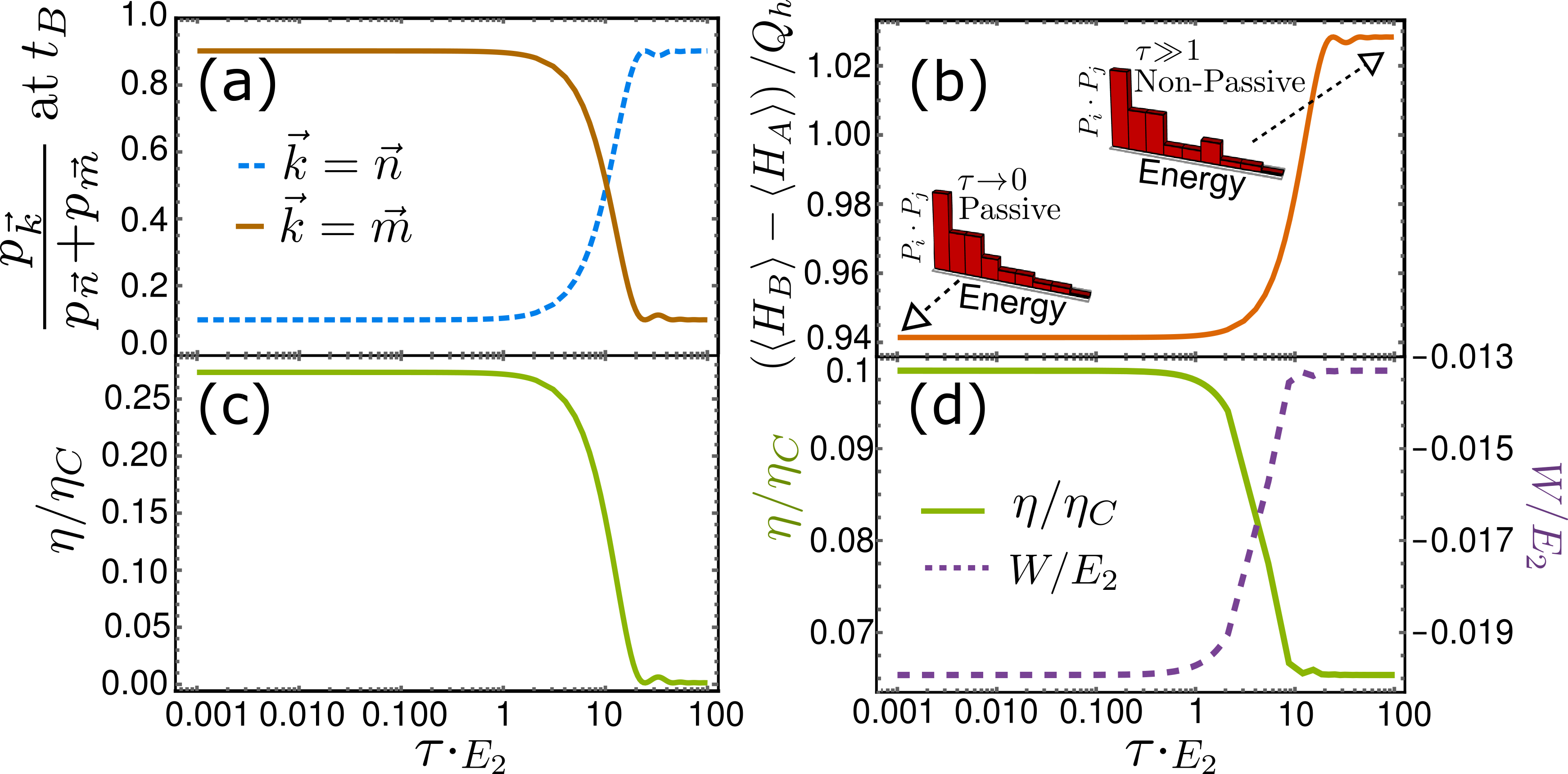}
	\caption{  (a) Relative occupation $p_{\vec{k}}$ of  levels $\vec{n} = \{1,1\}$ and $\vec{m} = \{0,2\}$ at point B, (b)  WF energy difference, (c)  and  efficiency of the whole cycle in units of Carnot efficiency as a function of $\tau$ for a WF formed by two copies. Inset (b) shows the level populations for the limits  $\tau \to 0$ (left dotted arrow) and $\tau \to \infty$ (right dotted arrow). Part (d) shows the efficiency and work extracted for a WF formed by three copies which has no classical counterpart.   Parameters: (a) - (c) $\beta_h E_2= 3.28,\beta_c E_2 = 6.66$, $E_0 = 0$, $E_1(0) = 1/3 E_2, \Delta E_1 = 1/3 E_2$, (d)
	$\beta_h E_2 = 1.09,\beta_c E_2 = 2.22$, $E_1(0) = 0.57 E_2, \Delta E_1 = 0.35 E_2$. }
	\label{fig:eficom}
\end{figure}

To compensate for this, in a second substroke, an interaction $f(t)$ between the QTs is activated after $t_{AB}$ for  a time duration 
$\tau = t_B - t_{AB}$. 
This interaction allows population transfer between energy levels 
$\ket{\vec{n}}$ and $\ket{\vec{m}}$, producing a collective passive state at the end of the
isentropic strokes for ideal swapping (see Fig.~\ref{fig:eficom}(a)-(b)). 
Such population exchanges decrease the total energy and the quantum relative entropy $\mathcal{D}$ in Eq.~\eqref{eq:full_efficiency}. 
The energy difference between these final states and the corresponding non-passive states  at the end of the 
QA processes (see Fig.~\ref{fig:fig1}(b)) is an extra supplementary work based on cooperative effects that can only be extracted in NQA strokes. 
Both cases 
are captured by the following choice for $f(t)$ during the isentropic compression
 \begin{align}
\label{eq:interaction}
f(t) &= \frac{\pi^2}{4 \tau} \sin\left[\pi (t - t_{AB})/\tau\right], \text{ for } t_{AB} \leq t \leq t_{B},
\end{align}    
where $f(t) = 0$ else and
$\int f(t) dt = \frac{\pi}{2}$. 
For $\tau \to 0$, the protocol~(\ref{eq:Ham_two_QT}) with $f(t)$ exactly implements the swap between $|\vec{n}\rangle$ and $|\vec{m}\rangle$ populations, creating a passive state at point $B$.
In contrast, for $\tau \to \infty$,~(\ref{eq:Ham_two_QT}) approaches a QA operation and populations stay constant. 
Then, since $f(t_A)=f(t_{AB}) = f(t_B) = 0$, the resulting state at point $B$ is non-passive.

To simulate the cycle, we solve the von-Neumann-Eq. with time-dependent $H_N(t)$ during $A$ to $B$ and with reversed protocol during $C$ to $D$. 
Figure~\ref{fig:eficom} shows exemplary results of our approach for different limits of $\tau$ for a WF made by two (a-c) and three (d) copies.
Panel (a) demonstrates the $\tau$ dependency of the level swap operation at point $B$. 
For $\tau \gg 1$ the final level occupation $|\vec{n}\rangle = |1,1\rangle$ and $|\vec{m}\rangle = |0,2\rangle$ does not change in comparison to the protocol without swap, such that due to the presence of level crossing the state is not passive any-more (see~\ref{fig:eficom}(b)-right inset). 
In contrast, reducing $\tau$ the occupations of the levels $|\vec{n}\rangle$ and $|\vec{m}\rangle$ start to swap, yielding a perfect swap for $\tau \to 0$. 
In this limit, the total energy   at point B is reduced and the state becomes passive (see Fig.~\ref{fig:eficom}(b)). 
Fig.~\ref{fig:eficom}(c) shows
the Otto cycle efficiency as function of  $\tau$.
For the chosen parameters, for  $\tau \gg 1$  the efficiency tends to the efficiency of the single copy engine or the two independent copies engine. 
In contrast, for  $\tau \to 0$,  cooperative effects between copies are included by our protocol, which perfects the swap quality,
exchanges the corresponding level populations, giving a boost to the extracted work and efficiency, which increases up to 0.25 $\eta_c$. 
This particular example could also be realized using a classical (continuous energy spectrum) WF.
Nevertheless, the example shown on figure~\ref{fig:eficom}(d) does not have a classical counterpart and requires a quantum WF (discrete energy spectrum):
In this last example, the WF is formed by three copies of the same system, such that its QA version for a WF composed of only single and double copy produces a passive states at points B and D. 
If the WF was classical, the three, four, etc. copies of the WF would result in a passive state and  there would not be any enhancement when going from the QA to the NQA limit. 
An efficiency increase starting for a  non-adiabatic three (or more) copies WF is therefore a signature of quantum behavior \cite{levy2018quantum,uzdin2015equivalence}.

{\bf Size scaling} -- In order to explore the collective behavior, we analyze how the extracted work and efficiency scale with the size of the WF, parametrized by the number of copies $N$. 
%
%
If the QTs act independently, i.e., non-cooperatively (ncp), the extracted work is just
$N$ times the work of a single (sgl) copy $W^{\rm ncp}=N W^{\rm sgl}$.
Because the heats have the same scaling ($Q_{h(c)}^{\rm ncp}=N Q_{h(c)}^{\rm sgl}$) the efficiency is independent of the system size $N$ in this case.
For our protocol, this corresponds to QA protocols, where cooperative effects are suppressed.
During QA cycles, $\rho_{B,N}$ and $\rho_{A,N}$ have the same populations given by the product state $\propto  \otimes_N \exp\left(-\beta_c H_{\rm QT}(t_A)\right)$ (as $f(t_A)=f(t_B)=0$, the energy eigenbases of $H_A$ and $H_B$ are identical for our example protocol). 
Also $\rho_D$ and $\rho_C$ have the same populations given by the product state $\propto \otimes_N \exp \left(-\beta_h H_{\rm QT}(t_B)\right)$ in QA cycles. 
This generates a linear -- non-cooperative -- energy scaling 
\begin{equation}
{\rm Tr}\lbrack H_{N}(t_\nu) \rho_{_{\nu,N}} \rbrack_{QA}  
	= N {\rm Tr}\lbrack H_{\rm QT}(t_{\nu}) \rho_{_{\nu,1}}\rbrack
\end{equation}
for $\nu \in \{A,B,C,D\}$.
According to Eq.~\eqref{eq:full_efficiency}, the efficiency would not change increasing $N$. 

But for the appropriate NQA modulation, cooperative effects produce passive collective states at $B$ and $D$.  
For finite-size passive states the energy scaling is sublinear~\cite{alicki2013entanglement}, i.e.,
for $\nu \in \{B,D\}$
\begin{equation}
{\rm Tr}\lbrack H_{N}(t_\nu) \bar{\rho}_{_{\nu,N}}\rbrack_{\rm NQA} \leq N {\rm Tr}\lbrack H_{\rm QT}(t_{\nu}) \bar{\rho}_{_{\nu,1}}\rbrack;\,.
\end{equation}
 \begin{figure}
	\includegraphics[width=0.49\textwidth]{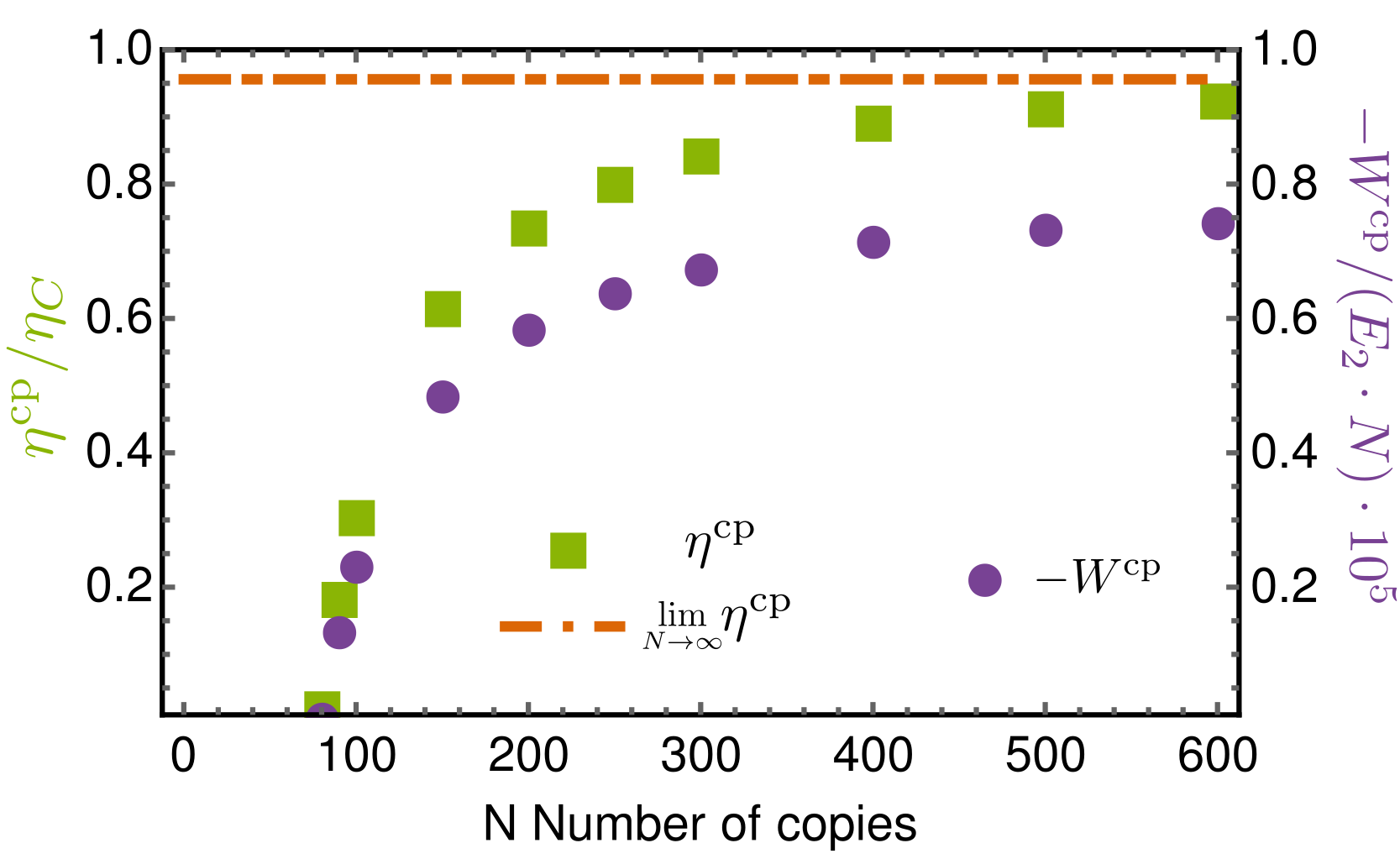}
	\caption{ Work (circles) and  normalized efficiency (squares) as function of the number of copies $N$ of the QT for a NQA heat engine with multiple perfect population swaps. The dot-dashed orange line shows the many-body efficiency limit which is asymptotically reached as the number of copies increases. Parameters: $\beta_h E_2 = 1.71, \beta_c E_2 = 1.85$, $E_0=0$, $E_1(0)=0.595 E_2$, $\Delta E_1=0.125 E_2$.
}
	\label{fig:multisystem_efficiency}
 \end{figure}
This produces the following behavior for the {\it cooperative} (cp) work and heats $W^{\rm cp}$, $Q^{\rm cp}_h$ and $Q^{\rm cp}_c$ as the WF size is increased: \textit{i)} the extracted work per copy increases, i.e., $W^{\rm cp}/N$ becomes more negative, so $|W^{\rm cp}|> N |W^{\rm sgl}|$; 
\textit{ii)} the incoming heat from the hot bath per copy increases, i.e., $Q^{\rm cp}_h/N$ becomes more positive, thus  $Q^{\rm cp}_h> N Q^{\rm sgl}_h$;
\textit{iii)} the heat flowing to the cold bath per copy decreases, i.e., $Q^{\rm cp}_c/N$  which is negative, tends to zero, so  $|Q^{\rm cp}_c|<N |Q^{\rm sgl}_c|$.
Finally, we obtain as consequence of \textit{ii} and \textit{iii} that the cooperative behaviour of the collective system enhances
the efficiency $\eta^{\rm cp}>\eta^{\rm ncp}$. 
The maximum efficiency is reached at large $N$, where a collective passive state tends to a thermal state~\cite{alicki2013entanglement} and 
$W^{\rm cp}$,  $Q^{\rm cp}_h$ and $Q^{\rm cp}_c$ become linear with the size of the WF.
As shown in Fig.~\ref{fig:multisystem_efficiency}(a), the efficiency converges to a many-body limit
(orange line). 
For perfect swaps, in the infinite-copy limit one has $\lim_{N \to \infty} D(\rho_{\rm D (B)}||\omega_{\rm D (B)})/Q_{\rm DA (BC)}^{(\omega)} \to 0$ in Eq.~\eqref{eq:full_efficiency}, 
and the efficiency converges to the many-body limit (see  SI~D)
\begin{equation}\label{eq:eta_macro}
\lim_{N\to \infty}\eta^{\rm cp} \to 1 + Q_{\rm c}^{(\omega)}/Q_{\rm h}^{(\omega)} \le \eta_C.
\end{equation}  

{\bf Conclusions} --
The lack of passivity of collective states represents a thermodynamic resource that, as we have shown, can be used to boost the work and efficiency of a heat engine when NQA operations are used during the isentropic strokes.   
Although we have used an external field to exploit the cooperative effects, the exchange of population among different components could be part of the intrinsic WF dynamics.
When the WF is a multipartite system, it is tempting to look for indications of entanglement~\cite{alicki2013entanglement} in the transformation from the non-passive to the passive state. 
Nevertheless, in general this is not required~\cite{hovhannisyan2013entanglement}, although quantum discords may be actually needed~ \cite{giorgi2015correlation}. 
In some cases,  the non-passivity of the collective state~\cite{gorecki1980passive} requires a quantum WF in order to boost the efficiency (see Fig, \ref{fig:eficom}d). 
Future research paths could explore the relation between quantum discords and the need of a quantum WF in order to boost the efficiency. 

\begin{acknowledgments}
The authors are indebted to Tobias Brandes who initiated the project. We thank Robert Alicki and Karen Hovhannisyan for useful discussions.
Financial support by the DFG (grant no. SFB 910, BR 1928/9-1) is gratefully acknowledged. DG-K's work was supported in part by the Center for Excitonics, an Energy Frontier Research Center funded by the US Department of Energy under award DE-SC0001088 (Energy conversion process)

\end{acknowledgments}

\appendix

\section{Carnot bound for perfect swaps}
\label{app:carnot1}
In this supplement we will prove that the efficiency of a quantum
heat engine that involves the population swaps is still limited by
the Carnot bound. Assume that $N$ energy levels contribute to the
work extraction,
\begin{gather}
W_{N}=\sum_{n,m}^{N}A_{nm};\nonumber \\
 A_{nm}=(E_{n}(t_{A})-E_{m}(t_{B}))(\frac{e^{-\beta_{h}E_{m}(t_{B})}}{Z_{h,t_{B}}}-\frac{e^{-\beta_{c}E_{n}(t_{A})}}{Z_{c,t_{A}}}).
\end{gather}

There are $N$ different $A_{nm}$ and the sum is over $n$ and $m$,
which can take any value from $1$ to $N$, as long as each energy
level, that is $E_{n}(t_{A})$ and $E_{m}(t_{B}),$ appears only once
in the sum. $Z_{i,t_{J}}=\sum_{n}e^{=\beta_{i}E_{n}(t_{J})}.$

An element with $n=m$ indicates an energy level without population
swap. If $n\neq m$, then the populations where swapped during the
isentropic stages. In a similar way, we write the heat coming from
the hot bath as

\begin{gather}
Q_{N,h}=\sum_{n,m}^{N}B_{nm}; \nonumber \\
B_{nm}=E_{m}(t_{B})(\frac{e^{-\beta_{h}E_{m}(t_{B})}}{Z_{h,t_{B}}}-\frac{e^{-\beta_{c}E_{n}(t_{A})}}{Z_{c,t_{A}}}).
\end{gather}

Notice that if $N=1$ the work and the heat is composed of a single
element, 

\begin{gather*}
W_{1}=A_{n_{0}m_{0}}= \nonumber \\
(E_{n_{0}}(t_{A})-E_{m_{0}}(t_{B}))(\frac{e^{-\beta_{h}E_{m_{0}}(t_{B})}}{Z_{h,t_{B}}}-\frac{e^{-\beta_{c}E_{n_{0}}(t_{A})}}{Z_{c,t_{A}}});\\
Q_{1h}=B_{n_{0}m_{0}}= \nonumber \\
(E_{m_{0}}(t_{B}))(\frac{e^{-\beta_{h}E_{m_{0}}(t_{B})}}{Z_{h,t_{B}}}-\frac{e^{-\beta_{c}E_{n_{0}}(t_{A})}}{Z_{c,t_{A}}}).
\end{gather*}

The condition for work extraction, e.g., $W_{1}\leq0$, is 

\[
1\geq\frac{E_{n_{0}}(t_{A})}{E_{m_{0}}(t_{B})}\geq\frac{\beta_{h}}{\beta_{c}}.
\]

Under this condition $Q_{1h}>0$ and the efficiency is limited by
the Carnot bound

\begin{equation}
\eta_{1}=\frac{-W_{1}}{Q_{1h}}=1-\frac{E_{n_{0}}(t_{A})}{E_{m_{0}}(t_{B})}<1-\frac{\beta_{h}}{\beta_{c}}=\eta_{C}.\label{eq:carone}
\end{equation}
Notice that this proof is independent if the levels where swapped
($n_{0}\neq m_{0})$ or not. 

Next, we will use induction to prove that in the case where $N$ levels
contribute to work extraction, also the efficiency, $\eta_{N}$, is
limited by the Carnot bound. Assume $\frac{-W_{N-1}}{Q_{N-1,h}}=\eta_{N-1}\leq\eta_{C}$,
now we calculate $W_{N}$ and $Q_{N,h}$, which we define as
\[
W_{N}=W_{N-1}+A_{n_{0}m_{0}};\quad Q_{N,h}=Q_{N-1}+B_{n_{0}m_{0}}.
\]

Now, we calculate if $\eta_{N}$ is larger than $\eta_{C}.$ For
this, we write $\eta_{N}$ as

\[
\eta_{N}=\frac{-W_{N-1}-A_{n_{0}m_{0}}}{Q_{N-1,h}+B_{n_{0}m_{0}}}=\eta_{C}+x
\]

and look what is required for having $x>0$.

\[
x=\frac{-W_{N-1}-A_{n_{0}m_{0}}-\eta_{C}(Q_{N-1,h}+B_{n_{0}m_{0}})}{Q_{N-1,h}+B_{n_{0}m_{0}}}.
\]

Using $-W_{N-1}-\eta_{C}Q_{N-1,h}\leq0$, we can bound $x,$

\[
x\leq\frac{-A_{n_{0}m_{0}}-\eta_{C}B_{n_{0}m_{0}}}{Q_{N-1,h}+B_{n_{0}m_{0}}}.
\]

By definition the incoming heat is positive, $Q_{N-1,h}+B_{n_{0}m_{0}}>0.$
Then, in order to violate the Carnot bound ($x>0)$ we need 

\begin{equation}
0<-A_{n_{0}m_{0}}-\eta_{C}B_{n_{0}m_{0}}\label{eq:ineq}
\end{equation}

There are two cases that we need to consider:
\begin{enumerate}
\item For $\frac{E_{n_{0}}(t_{A})}{E_{m_{0}}(t_{B})}\geq\frac{\beta_{h}}{\beta_{c}}\Rightarrow B_{n_{0}m_{0}}>0,$
and Eq.~\ref{eq:ineq} can be rewritten as 
\[
\eta_{C}<\frac{-A_{n_{0}m_{0}}}{B_{n_{0}m_{0}}}=1-\frac{E_{n_{0}}(t_{A})}{E_{m_{0}}(t_{B})}
\]
which is a contradiction because $\frac{E_{n_{0}}(t_{A})}{E_{m_{0}}(t_{B})}\geq\frac{\beta_{h}}{\beta_{c}}.$ 
\item For $\frac{E_{n_{0}}(t_{A})}{E_{m_{0}}(t_{B})}\leq\frac{\beta_{h}}{\beta_{c}}\Rightarrow B_{n_{0}m_{0}}\leq0,$
Eq.~\ref{eq:ineq} can be rewritten as
\[
\eta_{C}>\frac{-A_{n_{0}m_{0}}}{B_{n_{0}m_{0}}}=1-\frac{E_{n_{0}}(t_{A})}{E_{m_{0}}(t_{B})}
\]
which is a contradiction because $\frac{E_{n_{0}}(t_{A})}{E_{m_{0}}(t_{B})}\leq\frac{\beta_{h}}{\beta_{c}}.$ 
\end{enumerate}
Therefore, we can conclude that $x<0$ and the heat engine efficiency
is limited by the Carnot bound.

\section{General Carnot bound}
\label{app:carnot2}

Denoting the complete time evolution operator transferring the system from $A$ to $B$ by $U$, we can express the 
work~(Eq. (1) of the main text) as 
\begin{align}
W &= \trace{(U^\dagger H_B U - H_A)\rho_A}\nn
&\qquad+ \trace{(U H_A U^\dagger - H_B)\rho_C}\,.
\end{align}
Likewise, the heats from the hot and cold reservoirs~ (Eq. (2) of the main text) become
\begin{align}
Q_h &= \trace{H_B (\rho_C - U \rho_A U^\dagger)}\,,\nn
Q_c &= \trace{H_A (\rho_A - U^\dagger \rho_C U)}\,.
\end{align}
By adding all these contributions, we obtain the first law of thermodynamics for a cyclic process
$W+Q_{h}+Q_{c}=0$.
Furthermore, the sole purpose of the whole construction is to extract work $W_{AB}+W_{CD}<0$ by utilizing the incoming heat from the hot reservoir
$Q_{h}>0$, such that we define an efficiency as
\begin{align}
\eta = \frac{-W}{Q_{h}}=1-\frac{-Q_{c}}{Q_{h}}\,.
\end{align}

To estimate the entropic balance, we first consider that trivially, during the isentropic strokes the von-Neumann entropy does not change
$\Delta S_{AB} = 0 = \Delta S_{CD}$\,.
In contrast, during the equilibration strokes, the WF von-Neumann entropy changes due to the influx of heat from the hot reservoir
or the outflux of heat to the cold reservoir
\begin{align}
\Delta S_{BC} &= S(\rho_C) - S(U \rho_A U^\dagger) = S(\rho_C)-S(\rho_A)\,,\nn
\Delta S_{DA} &= S(\rho_A) - S(U^\dagger \rho_C U) = S(\rho_A)-S(\rho_C)\,.
\end{align}
However, the second law of thermodynamics now dictates that the change of WF entropy is bounded 
during the equilibration strokes
\begin{align}\label{EQ:entropy_bound}
\Delta S_{BC} - \beta_h Q_{h} \ge 0\,,\qquad
\Delta S_{DA} - \beta_c Q_{c} \ge 0\,.
\end{align}
This bounds the slopes of the cycle during the equilibration strokes in Fig. 1(b) of the main text, denoted by bold angles there.
These inequalities can be explicitly shown for quite general maps (Lindblad-Davies maps~\cite{spohn1978b}) 
but also general unitary evolutions~\cite{esposito2010b}.

Now, the condition that for cyclic operation, the total WF entropy change must cancel
$\Delta S_{BC}+\Delta S_{DA}=0$,
relates the heats with each other.
From $Q_c \le \beta_c^{-1} \Delta S_{DA} = -\beta_c^{-1} \Delta S_{BC}$ we can immediately conclude that
\begin{align}
-Q_c &\ge \beta_c^{-1} \Delta S_{BC} = \frac{\beta_h}{\beta_c} \beta_h^{-1} \Delta S_{BC} \ge \frac{\beta_h}{\beta_c} Q_h\,.
\end{align}
The cycle efficiency can therefore be universally (i.e., protocol-independent) bounded by Carnot efficiency
\begin{align}\label{EQ:carnot_bound}
\eta &\le 1 - \frac{\beta_h}{\beta_c} = 1-\frac{T_c}{T_h}\,.
\end{align}
This efficiency is hardly ever reached in actual scenarios.

\section{Generalized efficiency expression}
\label{app:eta}

For general density matrices $\rho_{\nu}$ at point $\nu \in \{{\rm A,B,C,D}\}$, see Fig.~1 in the main text, 
and Hamiltonians $H_\nu = H_N(t_\nu)$, the heat exchange $Q_{c}$ and $Q_{h}$ from Eq. 2 of the main text) can be re-expressed in terms of the 
quantum relative entropy between the actual state $\rho$ and thermal reference states 
\begin{equation}
\label{eq:def_thermal_state}
\omega_{\nu} = \frac{\exp(-\beta_\nu^{(\omega)} H_\nu)}{\trace{\exp(-\beta_\nu^{(\omega)} H_\nu)}}\,.
\end{equation} 
Here, the reference temperature $\beta_\nu^{(\omega)}$ is unambiguously fixed via the (nonlinear) condition 
$S(\rho_\nu) = S(\omega_{\nu})$ with $S(\rho)=-\trace{\rho \ln \rho}$ denoting 
the von-Neumann entropy of $\rho$.
With this definition, we see that the standard quantum relative entropy can be expressed by energy differences between the actual state and the 
reference state
\begin{align}
\label{eq:quant_rel_entropy}
\mathcal{D}(\rho_{ \nu}||\omega_{ \nu}) &= \trace{\rho_\nu \ln \rho_\nu - \rho_\nu \ln \omega_\nu}\nn
&=\beta_\nu^{(\omega)} [{\rm Tr}(H_\nu \rho_{\nu}) - {\rm Tr}(H_\nu \omega_{\nu})]\,.
\end{align}  
The heat exchange with the reservoirs can then be re-expressed using the corresponding reference states at each point of the protocol
\begin{align}
\label{eq:heat_eq_with_D}
Q_{h} &= \trace{H_B \omega_{\rm C}} -  \trace{H_B \omega_{\rm B}} \nn
	&\quad - \trace{H_B (\omega_{\rm C} - \rho_C)} + \trace{H_B (\omega_{\rm B} - \rho_B)} \nn
	& = Q_{\rm h}^{(\omega)} - (\beta_{\rm B}^{(\omega)})^{-1} \mathcal{D}(\rho_B,\omega_{\rm B}) + (\beta_{\rm C}^{(\omega)})^{-1} \mathcal{D}(\rho_C,\omega_{\rm C}),\\
Q_{c}& = Q_{\rm c}^{(\omega)} + (\beta_{\rm A}^{(\omega)})^{-1} \mathcal{D}(\rho_A,\omega_{\rm A}) - (\beta_{\rm D}^{(\omega)})^{-1} \mathcal{D}(\rho_D,\omega_{\rm D}).
\end{align}
Here,  $Q_{h(c)}^{(\omega)} =  \trace{H_{B(A)} \omega_{C(A)}} -  \trace{H_{B(A)} \omega_{B(D)}} $ denote the heat exchange with hot (cold) baths for thermal reference states, compare the axes in Fig.~1(b) in the main text.
With Eq.~\eqref{eq:heat_eq_with_D} the efficiency $\eta$ yields
\begin{widetext}
\begin{equation}
\eta  = 1 - \frac{-Q_{\rm c}}{Q_{\rm h}} \\
	= 1 - \left[1 + 
			\frac{\mathcal{D}(\rho_D||\omega_{\rm D})}
				{\beta_{\rm D}^{(\omega)} ( - Q_{\rm c}^{(\omega)})} 
			- \frac{\mathcal{D}(\rho_A||\omega_{\rm A})}
			{\beta_{\rm A}^{(\omega)} (- Q_{\rm c}^{(\omega)})}
			\right] \cdot 
			\frac{1}
			{1 - 
				\frac{\mathcal{D}(\rho_B||\omega_{\rm B})}
				{\beta_{\rm B}^{(\omega)} ( - Q_{\rm h}^{(\omega)})}  + \frac{\mathcal{D}(\rho_C||\omega_{\rm C})}
				{\beta_{\rm C}^{(\omega)} ( - Q_{\rm h}^{(\omega)})}} 
			\cdot 
			\frac{- Q_{\rm c}^{(\omega)}}
				{Q_{\rm h}^{(\omega)}}.
\end{equation}
\end{widetext}
%
 
In the main text we assume that $\rho_A$ and $\rho_C$ are thermal distributions, 
thus they would coincide with the corresponding thermal reference states, i.e. $\rho_A = \omega_{\rm A}$ and $\rho_C = \omega_{\rm C}$, 
and the corresponding relative entropies vanish. 
The equation above then reduces to Eq.~(3) in the main text. 

If the cycle is actually performed with the thermal reference states, the second-law inequalities~(\ref{EQ:entropy_bound}) still hold with
$Q_{c/h}\to Q_{c/h}^{(\omega)}$ and temperatures as well as entropy differences unchanged, which can be used to show that also in this limit the efficiency is bound by Carnot efficiency.
In particular, the slopes of the dashed lines in Fig.~1(b) in the main text are given by
\begin{align}
\frac{\Delta S_{\rm BC}}{Q_h^{(w)}} &= \frac{\Delta S_{\rm BC}}{Q_h + [\beta_B^{(\omega)}]^{-1} \mathcal{D}(\rho_B || \omega_B)}\,,\nn
\abs{\frac{\Delta S_{\rm DA}}{Q_c^{(w)}}} &= \abs{\frac{\Delta S_{\rm DA}}{Q_c + [\beta_D^{(\omega)}]^{-1} \mathcal{D}(\rho_D || \omega_D)}}\,,
\end{align}
which shows that the slope of the hot reference equilibration stroke decreases, whereas (due to $Q_c<0$) the slope of the cold reference equilibration stroke increases in comparison to the original slope (dashed lines versus solid lines in Fig.~1(b) of the main text).
Most important however, since the thermal reference states have the minimum energy for a given entropy and Hamiltonian, their Otto cycle curve is extremal, and no other isentropic process with the same Hamiltonians can have a more optimal heat balance.

\section{Many-body limit of efficiency $\eta^{cp}$}
\label{app:ninfty_proof}

In the following we show that Eq.~3 in the main text converges for our protocol to a simple limit in case of infinitely many copies. 
For simplicity, we also assume a perfect swap after the level ramp, such that $\rho_B \to \bar{\rho}_B$ and $\rho_D \to \bar{\rho}_D$. 
The swap exchanges the populations of all levels with crossing and will be represented by the unitary $U$
(note the difference to Sec.~\ref{app:carnot2}).
Doing so, the system state just after the level ramp (first substroke) $\rho_{AB} = \otimes_{j=1}^N \rho_{A,1}^{(j)}$ is non-passive, 
where $\rho_{A,1}$ denotes the single qutrit density matrix at point $A$ of the Otto cycle.
After the swap, it becomes passive 
$\bar{\rho}_{B} = U \rho_{AB} U^\dag$. 
Further, at point $B$ a thermal reference state $\omega_{B}$ (of $N$ copies) can be defined with reference temperature
$\beta_B^{(\omega)}$ as explained in Eq.~(\ref{eq:def_thermal_state}).
Since the Hamiltonian $H_B$ is additive due to $f(t_B)=0$, this thermal reference state is just given by the tensor product
of the single-qutrit reference states.
Hence, the conditions in the work by R. Alicki and M. Fannes~\cite{alicki2013entanglement} are met. 

In their notation, $\sigma_{\otimes^N \rho}$ denotes a passive state of $N$ copies which was created by a unitary from a density matrix 
$\rho$, and $\omega_{\bar \beta}$ denotes the (single-copy) thermal reference state of $\rho$ with reference temperature $\bar\beta$. 
From Eqns.~(16) and~(17) of their work we obtain that -- 
for a non-interacting many-body Hamiltonian composed of $N$ identical contributions 
$H_1$ each, i.e., $H^{(N)} = \sum_{j=1}^N H_1^{(j)}$ -- one can write
\begin{align}
\trace{\sigma_{\otimes^n \rho} H^{(N)}} \stackrel{N \to \infty}{\to} N \trace{\omega_{\bar \beta} H_1} + o(N)\,,
\end{align}
where we employ the notation $o(x)$ to express that the remainder term grows slower than $x$, i.e., 
$\lim_{N\to\infty} \frac{o(N)}{N} = 0$.
Thus, the value for the energy in a passive state approaches the energy in the corresponding thermal reference state. 
Alternatively, this can be also seen combining Eqns.~(20) and~(21) of Ref.~\cite{alicki2013entanglement}.
Applying this result to our notation we get consequently, 
\begin{align}
\frac{\mathcal{D}(\bar{\rho}_{{B}}||\omega_{{B}})}
{\beta_{\rm B}^{(\omega)}} &= \trace{\bar{\rho}_{{B}} H_B} - \trace{\omega_{{B}} H_B}\nn
&= \trace{U \rho_{{AB}} U^\dag  H_B} - \trace{\omega_{{B}} H_B}\nn
&\to N \cdot \trace{\omega_{{AB},1}  H_{B,1}} - N \cdot \trace{\omega_{{B},1} H_{B,1}}\nn
&\qquad + o(N)\nn
&= o(N)\,,
\end{align}
where in the last step we have used that as before and after the swap both 
entropy and Hamiltonian are the same, the thermal reference states are equal:
As the Hamiltonian at $t_{AB}$ and $t_B$ is non-interacting, they are thus tensor products of
identical single-copy states, i.e., $\omega_{{AB},1} = \omega_{{B},1}$. 
Also, by construction, the quantity $Q_{\rm BC}^{(\omega)}$ is extensive, since $H_B$ is additive, 
such that both the thermal reference state $\omega_B$ and final state $\rho_C$ are product states.
Thus, we obtain
\begin{equation}
	\frac{\mathcal{D}(\bar{\rho}_B||\omega_{\rm B})}
{\beta_{\rm B}^{(\omega)} ( - Q_{ \rm BC}^{(\omega)})} \stackrel{N \to \infty}{\to } 0\,,  
\end{equation}
and using similar arguments for the cold reservoir
\begin{equation}
\frac{\mathcal{D}(\bar{\rho}_D||\omega_{\rm D})}
{\beta_{\rm D}^{(\omega)} ( - Q_{\rm DA}^{(\omega)})} \stackrel{N \to \infty}{\to } 0\,.
\end{equation}
In total, Eq. 3 in the main text simplifies in the many-body limit to
\begin{equation}
\lim_{N\to \infty}\eta^{\rm cp} = 1 - \frac{- Q_{\rm DA}^{(\omega)}}
							{Q_{\rm BC}^{(\omega)}}\,,
\end{equation}
which corresponds to a system which reaches at each point in the protocol the thermal reference state. 
Naturally, this tightens the Carnot bound~(\ref{EQ:carnot_bound}) and shows that the thermal reference states
are a useful concept.

\section{Qubit implementation}

For a number of qubits described by Pauli matrices $\sigma_\alpha^i$ with $\alpha\in\{x,y,z\}$ and $i\in\{1,2,\ldots\}$, we 
define the large-spin operators $J_\alpha = 1/2 \sum_i \sigma_\alpha^i$ and
the corresponding ladder operators $J_\pm = \sum_i \sigma_\pm^i$.
The first is just the $z$ component of the total angular momentum operator, and one can directly check that $[J_-, J_+] = -2 J_z$.
Furthermore, we introduce the angular momentum eigenstates as $J_z \ket{m} = m \ket{m}$.
Such large spin operators may arise in the fully symmetric subspace of two qubits, which naturally implements a qutrit.
In this case we have
$J_z = (\sigma^z_1 + \sigma^z_2)/2$ and $J_+ = \sigma^+_1 + \sigma^+_2$ with $\sigma^+_i =(\sigma^x_i + \ii \sigma^y_i)/2$.
Next, we note that the time-dependent Hamiltonian
\begin{align}
H_0(t) = \Omega \left[J_z + b(t) \left(J_z^2 - \f{1}\right)\right] 
= \left(\begin{array}{ccc}
-\Omega & 0 & 0\\
0 & -b(t) \Omega & 0\\
0 & 0 & +\Omega
\end{array}\right)
\end{align}
only moves the $m=0$ level and secondly, commutes with itself at different times, as exemplified by Eq.~(4) in the main text.
This implies that the corresponding time evolution operator
\begin{align}
U(t) &= \hat{\tau} \exp\left\{-\ii \int_0^t H_0(t') dt'\right\}\nn
&= \left(\begin{array}{ccc}
+\exp(\ii \Omega t) & 0 & 0\\
0 & \exp(\ii \Omega \int_0^t b(t') dt') & 0\\
0 & 0 & \exp(-\ii \Omega t)
\end{array}\right)
\end{align}
is diagonal and thereby does not induce transitions between the time-independent eigenstates, no matter how fast $b(t)$ changes, i.e., 
it is always a QA protocol.
The only effect of this ramp is thus a change of the system energy while the middle level is moved.

We now apply such a quench of the intermediate level to two qutrits (denoted by $1$ and $2$), initially prepared in identical thermal states.
\begin{align}
H_0(t) &= \Omega \left[J_{z,1} + b(t) \left((J_{z,1})^2 - \f{1}\right)\right]\nn
&\qquad+ \Omega \left[J_{z,2} + b(t) \left((J_{z,2})^2 - \f{1}\right)\right]\,.
\end{align}
Afterwards, the resulting state need not be passive anymore, since energies between the subsystems are additive but populations are multiplicative.
The desired swap operation $S$, which exchanges $\ket{m_1=0}\bra{m_1=0} \otimes \ket{m_2=0}\bra{m_2=0}$ with $\ket{m_1=-1}\bra{m_1=-1} \otimes \ket{m_2=+1}\bra{m_2=+1}$  
can now be implemented by a unitary operation of the form
\begin{align}
S &= e^{-\ii \pi H_{\rm sw}}\,,\nn
H_{\rm sw}&=\frac{1}{4} \left[J_-(\f{1}-J_z^2)\right]_1 \otimes \left[J_+(\f{1}-J_z^2)\right]_2\nn
&\qquad+ \frac{1}{4} \left[(\f{1}-J_z^2) J_+\right]_1 \otimes \left[(\f{1}-J_z^2)J_-\right]_2\nn
&\qquad- \frac{1}{2} \left[\f{1}-J_z^2\right]_1 \otimes \left[\f{1}-J_z^2\right]_2\nn
&\qquad- \frac{1}{8} \left[J_z^2 - J_z\right]_1 \otimes \left[J_z^2+J_z\right]_2\,.
\end{align}
We note that when e.g. the qutrit is implemented in the symmetric subspace of two physical qubits, the swap will act non-trivially on the other
subspaces of vanishing total angular momentum. 
This means that to make the protocol work with qubits, they should only be prepared in the fully symmetric subspace.
Alternatively, a penalty Hamiltonian could be used to separate the undesired subspaces energetically.


\begin{thebibliography}{33}
\expandafter\ifx\csname natexlab\endcsname\relax\def\natexlab#1{#1}\fi
\expandafter\ifx\csname bibnamefont\endcsname\relax
  \def\bibnamefont#1{#1}\fi
\expandafter\ifx\csname bibfnamefont\endcsname\relax
  \def\bibfnamefont#1{#1}\fi
\expandafter\ifx\csname citenamefont\endcsname\relax
  \def\citenamefont#1{#1}\fi
\expandafter\ifx\csname url\endcsname\relax
  \def\url#1{\texttt{#1}}\fi
\expandafter\ifx\csname urlprefix\endcsname\relax\def\urlprefix{URL }\fi
\providecommand{\bibinfo}[2]{#2}
\providecommand{\eprint}[2][]{\url{#2}}

\bibitem[{\citenamefont{Landau et~al.}(1980)\citenamefont{Landau, Lifshitz, and
  Pitaevskii}}]{landau1980statistical}
\bibinfo{author}{\bibfnamefont{L.~D.} \bibnamefont{Landau}},
  \bibinfo{author}{\bibfnamefont{E.~M.} \bibnamefont{Lifshitz}},
  \bibnamefont{and}
  \bibinfo{author}{\bibfnamefont{L.}~\bibnamefont{Pitaevskii}},
  \emph{\bibinfo{title}{Statistical physics, part i}} (\bibinfo{year}{1980}).

\bibitem[{\citenamefont{Kosloff}(2013)}]{kosloff2013quantum}
\bibinfo{author}{\bibfnamefont{R.}~\bibnamefont{Kosloff}},
  \bibinfo{journal}{Entropy} \textbf{\bibinfo{volume}{15}},
  \bibinfo{pages}{2100} (\bibinfo{year}{2013}).

\bibitem[{\citenamefont{Gelbwaser-Klimovsky
  et~al.}(2015)\citenamefont{Gelbwaser-Klimovsky, Niedenzu, and
  Kurizki}}]{gelbwaser2015thermodynamics}
\bibinfo{author}{\bibfnamefont{D.}~\bibnamefont{Gelbwaser-Klimovsky}},
  \bibinfo{author}{\bibfnamefont{W.}~\bibnamefont{Niedenzu}}, \bibnamefont{and}
  \bibinfo{author}{\bibfnamefont{G.}~\bibnamefont{Kurizki}}, in
  \emph{\bibinfo{booktitle}{Advances In Atomic, Molecular, and Optical
  Physics}} (\bibinfo{publisher}{Elsevier}, \bibinfo{year}{2015}),
  vol.~\bibinfo{volume}{64}, pp. \bibinfo{pages}{329--407}.

\bibitem[{\citenamefont{Allahverdyan et~al.}(2008)\citenamefont{Allahverdyan,
  Johal, and Mahler}}]{allahverdyan2008work}
\bibinfo{author}{\bibfnamefont{A.~E.} \bibnamefont{Allahverdyan}},
  \bibinfo{author}{\bibfnamefont{R.~S.} \bibnamefont{Johal}}, \bibnamefont{and}
  \bibinfo{author}{\bibfnamefont{G.}~\bibnamefont{Mahler}},
  \bibinfo{journal}{Physical Review E} \textbf{\bibinfo{volume}{77}},
  \bibinfo{pages}{041118} (\bibinfo{year}{2008}).

\bibitem[{\citenamefont{Berry}(1984)}]{berry1984a}
\bibinfo{author}{\bibfnamefont{M.~V.} \bibnamefont{Berry}},
  \bibinfo{journal}{Proceedings of the Royal Society of London Series A}
  \textbf{\bibinfo{volume}{392}}, \bibinfo{pages}{45} (\bibinfo{year}{1984}),
  \urlprefix\url{https://www.jstor.org/stable/2397741}.

\bibitem[{\citenamefont{Wu}(1989)}]{wu1989a}
\bibinfo{author}{\bibfnamefont{Z.}~\bibnamefont{Wu}},
  \bibinfo{journal}{Physical Review A} \textbf{\bibinfo{volume}{40}},
  \bibinfo{pages}{2184} (\bibinfo{year}{1989}).

\bibitem[{\citenamefont{Sarandy et~al.}(2004)\citenamefont{Sarandy, Wu, and
  Lidar}}]{sarandy2004a}
\bibinfo{author}{\bibfnamefont{M.~S.} \bibnamefont{Sarandy}},
  \bibinfo{author}{\bibfnamefont{L.-A.} \bibnamefont{Wu}}, \bibnamefont{and}
  \bibinfo{author}{\bibfnamefont{D.~A.} \bibnamefont{Lidar}},
  \bibinfo{journal}{Quantum Information Processing}
  \textbf{\bibinfo{volume}{3}}, \bibinfo{pages}{331} (\bibinfo{year}{2004}).

\bibitem[{\citenamefont{Schaller et~al.}(2006)\citenamefont{Schaller, Mostame,
  and Sch\"utzhold}}]{schaller2006b}
\bibinfo{author}{\bibfnamefont{G.}~\bibnamefont{Schaller}},
  \bibinfo{author}{\bibfnamefont{S.}~\bibnamefont{Mostame}}, \bibnamefont{and}
  \bibinfo{author}{\bibfnamefont{R.}~\bibnamefont{Sch\"utzhold}},
  \bibinfo{journal}{Physical Review A} \textbf{\bibinfo{volume}{73}},
  \bibinfo{pages}{062307} (\bibinfo{year}{2006}).

\bibitem[{\citenamefont{Jansen et~al.}(2007)\citenamefont{Jansen, Ruskai, and
  Seiler}}]{jansen2007a}
\bibinfo{author}{\bibfnamefont{S.}~\bibnamefont{Jansen}},
  \bibinfo{author}{\bibfnamefont{M.~B.} \bibnamefont{Ruskai}},
  \bibnamefont{and} \bibinfo{author}{\bibfnamefont{R.}~\bibnamefont{Seiler}},
  \bibinfo{journal}{Journal of Mathematical Physics}
  \textbf{\bibinfo{volume}{48}}, \bibinfo{pages}{102111}
  (\bibinfo{year}{2007}).

\bibitem[{\citenamefont{Zheng and Poletti}(2014)}]{zheng2014work}
\bibinfo{author}{\bibfnamefont{Y.}~\bibnamefont{Zheng}} \bibnamefont{and}
  \bibinfo{author}{\bibfnamefont{D.}~\bibnamefont{Poletti}},
  \bibinfo{journal}{Physical Review E} \textbf{\bibinfo{volume}{90}},
  \bibinfo{pages}{012145} (\bibinfo{year}{2014}).

\bibitem[{\citenamefont{Quan et~al.}(2007)\citenamefont{Quan, Liu, Sun, and
  Nori}}]{quan2007quantum}
\bibinfo{author}{\bibfnamefont{H.}~\bibnamefont{Quan}},
  \bibinfo{author}{\bibfnamefont{Y.-x.} \bibnamefont{Liu}},
  \bibinfo{author}{\bibfnamefont{C.}~\bibnamefont{Sun}}, \bibnamefont{and}
  \bibinfo{author}{\bibfnamefont{F.}~\bibnamefont{Nori}},
  \bibinfo{journal}{Physical Review E} \textbf{\bibinfo{volume}{76}},
  \bibinfo{pages}{031105} (\bibinfo{year}{2007}).

\bibitem[{\citenamefont{Jaramillo et~al.}(2016)\citenamefont{Jaramillo, Beau,
  and del Campo}}]{jaramillo2016quantum}
\bibinfo{author}{\bibfnamefont{J.}~\bibnamefont{Jaramillo}},
  \bibinfo{author}{\bibfnamefont{M.}~\bibnamefont{Beau}}, \bibnamefont{and}
  \bibinfo{author}{\bibfnamefont{A.}~\bibnamefont{del Campo}},
  \bibinfo{journal}{New Journal of Physics} \textbf{\bibinfo{volume}{18}},
  \bibinfo{pages}{075019} (\bibinfo{year}{2016}).

\bibitem[{\citenamefont{Hardal et~al.}(2018)\citenamefont{Hardal, Paternostro,
  and M{\"u}stecapl{\i}o{\u{g}}lu}}]{hardal2018phase}
\bibinfo{author}{\bibfnamefont{A.~{\"U}.} \bibnamefont{Hardal}},
  \bibinfo{author}{\bibfnamefont{M.}~\bibnamefont{Paternostro}},
  \bibnamefont{and} \bibinfo{author}{\bibfnamefont{{\"O}.~E.}
  \bibnamefont{M{\"u}stecapl{\i}o{\u{g}}lu}}, \bibinfo{journal}{Physical Review
  E} \textbf{\bibinfo{volume}{97}}, \bibinfo{pages}{042127}
  (\bibinfo{year}{2018}).

\bibitem[{\citenamefont{Niedenzu and Kurizki}(2018)}]{niedenzu2018cooperative}
\bibinfo{author}{\bibfnamefont{W.}~\bibnamefont{Niedenzu}} \bibnamefont{and}
  \bibinfo{author}{\bibfnamefont{G.}~\bibnamefont{Kurizki}},
  \bibinfo{journal}{arXiv preprint arXiv:1806.10810}  (\bibinfo{year}{2018}).

\bibitem[{\citenamefont{Niedenzu et~al.}(2015)\citenamefont{Niedenzu,
  Gelbwaser-Klimovsky, and Kurizki}}]{niedenzu2015performance}
\bibinfo{author}{\bibfnamefont{W.}~\bibnamefont{Niedenzu}},
  \bibinfo{author}{\bibfnamefont{D.}~\bibnamefont{Gelbwaser-Klimovsky}},
  \bibnamefont{and} \bibinfo{author}{\bibfnamefont{G.}~\bibnamefont{Kurizki}},
  \bibinfo{journal}{Physical Review E} \textbf{\bibinfo{volume}{92}},
  \bibinfo{pages}{042123} (\bibinfo{year}{2015}).

\bibitem[{\citenamefont{Del~Campo et~al.}(2014)\citenamefont{Del~Campo, Goold,
  and Paternostro}}]{del2014more}
\bibinfo{author}{\bibfnamefont{A.}~\bibnamefont{Del~Campo}},
  \bibinfo{author}{\bibfnamefont{J.}~\bibnamefont{Goold}}, \bibnamefont{and}
  \bibinfo{author}{\bibfnamefont{M.}~\bibnamefont{Paternostro}},
  \bibinfo{journal}{Scientific reports} \textbf{\bibinfo{volume}{4}},
  \bibinfo{pages}{6208} (\bibinfo{year}{2014}).

\bibitem[{\citenamefont{Halpern et~al.}(2017)\citenamefont{Halpern, White,
  Gopalakrishnan, and Refael}}]{halpern2017mbl}
\bibinfo{author}{\bibfnamefont{N.~Y.} \bibnamefont{Halpern}},
  \bibinfo{author}{\bibfnamefont{C.~D.} \bibnamefont{White}},
  \bibinfo{author}{\bibfnamefont{S.}~\bibnamefont{Gopalakrishnan}},
  \bibnamefont{and} \bibinfo{author}{\bibfnamefont{G.}~\bibnamefont{Refael}},
  \bibinfo{journal}{arXiv preprint arXiv:1707.07008}  (\bibinfo{year}{2017}).

\bibitem[{\citenamefont{Gelbwaser-Klimovsky
  et~al.}(2018)\citenamefont{Gelbwaser-Klimovsky, Bylinskii, Gangloff, Islam,
  Aspuru-Guzik, and Vuletic}}]{gelbwaser2018single}
\bibinfo{author}{\bibfnamefont{D.}~\bibnamefont{Gelbwaser-Klimovsky}},
  \bibinfo{author}{\bibfnamefont{A.}~\bibnamefont{Bylinskii}},
  \bibinfo{author}{\bibfnamefont{D.}~\bibnamefont{Gangloff}},
  \bibinfo{author}{\bibfnamefont{R.}~\bibnamefont{Islam}},
  \bibinfo{author}{\bibfnamefont{A.}~\bibnamefont{Aspuru-Guzik}},
  \bibnamefont{and} \bibinfo{author}{\bibfnamefont{V.}~\bibnamefont{Vuletic}},
  \bibinfo{journal}{Physical review letters} \textbf{\bibinfo{volume}{120}},
  \bibinfo{pages}{170601} (\bibinfo{year}{2018}).

\bibitem[{\citenamefont{Levy and Gelbwaser-Klimovsky}(2018)}]{levy2018quantum}
\bibinfo{author}{\bibfnamefont{A.}~\bibnamefont{Levy}} \bibnamefont{and}
  \bibinfo{author}{\bibfnamefont{D.}~\bibnamefont{Gelbwaser-Klimovsky}},
  \bibinfo{journal}{arXiv preprint arXiv:1803.05586}  (\bibinfo{year}{2018}).

\bibitem[{\citenamefont{Allahverdyan et~al.}(2004)\citenamefont{Allahverdyan,
  Balian, and Nieuwenhuizen}}]{allahverdyan2004maximal}
\bibinfo{author}{\bibfnamefont{A.~E.} \bibnamefont{Allahverdyan}},
  \bibinfo{author}{\bibfnamefont{R.}~\bibnamefont{Balian}}, \bibnamefont{and}
  \bibinfo{author}{\bibfnamefont{T.~M.} \bibnamefont{Nieuwenhuizen}},
  \bibinfo{journal}{EPL (Europhysics Letters)} \textbf{\bibinfo{volume}{67}},
  \bibinfo{pages}{565} (\bibinfo{year}{2004}).

\bibitem[{\citenamefont{Niedenzu et~al.}(2016)\citenamefont{Niedenzu,
  Gelbwaser-Klimovsky, Kofman, and Kurizki}}]{niedenzu2016operation}
\bibinfo{author}{\bibfnamefont{W.}~\bibnamefont{Niedenzu}},
  \bibinfo{author}{\bibfnamefont{D.}~\bibnamefont{Gelbwaser-Klimovsky}},
  \bibinfo{author}{\bibfnamefont{A.~G.} \bibnamefont{Kofman}},
  \bibnamefont{and} \bibinfo{author}{\bibfnamefont{G.}~\bibnamefont{Kurizki}},
  \bibinfo{journal}{New Journal of Physics} \textbf{\bibinfo{volume}{18}},
  \bibinfo{pages}{083012} (\bibinfo{year}{2016}).

\bibitem[{\citenamefont{Pusz and Woronowicz}(1978)}]{pusz1978passive}
\bibinfo{author}{\bibfnamefont{W.}~\bibnamefont{Pusz}} \bibnamefont{and}
  \bibinfo{author}{\bibfnamefont{S.}~\bibnamefont{Woronowicz}},
  \bibinfo{journal}{Communications in Mathematical Physics}
  \textbf{\bibinfo{volume}{58}}, \bibinfo{pages}{273} (\bibinfo{year}{1978}).

\bibitem[{\citenamefont{Lenard}(1978)}]{lenard1978thermodynamical}
\bibinfo{author}{\bibfnamefont{A.}~\bibnamefont{Lenard}},
  \bibinfo{journal}{Journal of Statistical Physics}
  \textbf{\bibinfo{volume}{19}}, \bibinfo{pages}{575} (\bibinfo{year}{1978}).

\bibitem[{\citenamefont{Gelbwaser-Klimovsky
  et~al.}(2013)\citenamefont{Gelbwaser-Klimovsky, Alicki, and
  Kurizki}}]{gelbwaser2013work}
\bibinfo{author}{\bibfnamefont{D.}~\bibnamefont{Gelbwaser-Klimovsky}},
  \bibinfo{author}{\bibfnamefont{R.}~\bibnamefont{Alicki}}, \bibnamefont{and}
  \bibinfo{author}{\bibfnamefont{G.}~\bibnamefont{Kurizki}},
  \bibinfo{journal}{EPL (Europhysics Letters)} \textbf{\bibinfo{volume}{103}},
  \bibinfo{pages}{60005} (\bibinfo{year}{2013}).

\bibitem[{\citenamefont{Gorecki and Pusz}(1980)}]{gorecki1980passive}
\bibinfo{author}{\bibfnamefont{J.}~\bibnamefont{Gorecki}} \bibnamefont{and}
  \bibinfo{author}{\bibfnamefont{W.}~\bibnamefont{Pusz}},
  \bibinfo{journal}{Letters in Mathematical Physics}
  \textbf{\bibinfo{volume}{4}}, \bibinfo{pages}{433} (\bibinfo{year}{1980}).

\bibitem[{\citenamefont{Ro{\ss}nagel et~al.}(2016)\citenamefont{Ro{\ss}nagel,
  Dawkins, Tolazzi, Abah, Lutz, Schmidt-Kaler, and
  Singer}}]{rossnagel2016single}
\bibinfo{author}{\bibfnamefont{J.}~\bibnamefont{Ro{\ss}nagel}},
  \bibinfo{author}{\bibfnamefont{S.~T.} \bibnamefont{Dawkins}},
  \bibinfo{author}{\bibfnamefont{K.~N.} \bibnamefont{Tolazzi}},
  \bibinfo{author}{\bibfnamefont{O.}~\bibnamefont{Abah}},
  \bibinfo{author}{\bibfnamefont{E.}~\bibnamefont{Lutz}},
  \bibinfo{author}{\bibfnamefont{F.}~\bibnamefont{Schmidt-Kaler}},
  \bibnamefont{and} \bibinfo{author}{\bibfnamefont{K.}~\bibnamefont{Singer}},
  \bibinfo{journal}{Science} \textbf{\bibinfo{volume}{352}},
  \bibinfo{pages}{325} (\bibinfo{year}{2016}).

\bibitem[{\citenamefont{Newman et~al.}(2017)\citenamefont{Newman, Mintert, and
  Nazir}}]{newman2017a}
\bibinfo{author}{\bibfnamefont{D.}~\bibnamefont{Newman}},
  \bibinfo{author}{\bibfnamefont{F.}~\bibnamefont{Mintert}}, \bibnamefont{and}
  \bibinfo{author}{\bibfnamefont{A.}~\bibnamefont{Nazir}},
  \bibinfo{journal}{Physical Review E} \textbf{\bibinfo{volume}{95}},
  \bibinfo{pages}{032139} (\bibinfo{year}{2017}).

\bibitem[{\citenamefont{Alicki and Fannes}(2013)}]{alicki2013entanglement}
\bibinfo{author}{\bibfnamefont{R.}~\bibnamefont{Alicki}} \bibnamefont{and}
  \bibinfo{author}{\bibfnamefont{M.}~\bibnamefont{Fannes}},
  \bibinfo{journal}{Physical Review E} \textbf{\bibinfo{volume}{87}},
  \bibinfo{pages}{042123} (\bibinfo{year}{2013}).

\bibitem[{\citenamefont{Uzdin et~al.}(2015)\citenamefont{Uzdin, Levy, and
  Kosloff}}]{uzdin2015equivalence}
\bibinfo{author}{\bibfnamefont{R.}~\bibnamefont{Uzdin}},
  \bibinfo{author}{\bibfnamefont{A.}~\bibnamefont{Levy}}, \bibnamefont{and}
  \bibinfo{author}{\bibfnamefont{R.}~\bibnamefont{Kosloff}},
  \bibinfo{journal}{Physical Review X} \textbf{\bibinfo{volume}{5}},
  \bibinfo{pages}{031044} (\bibinfo{year}{2015}).

\bibitem[{\citenamefont{Hovhannisyan et~al.}(2013)\citenamefont{Hovhannisyan,
  Perarnau-Llobet, Huber, and Ac{\'\i}n}}]{hovhannisyan2013entanglement}
\bibinfo{author}{\bibfnamefont{K.~V.} \bibnamefont{Hovhannisyan}},
  \bibinfo{author}{\bibfnamefont{M.}~\bibnamefont{Perarnau-Llobet}},
  \bibinfo{author}{\bibfnamefont{M.}~\bibnamefont{Huber}}, \bibnamefont{and}
  \bibinfo{author}{\bibfnamefont{A.}~\bibnamefont{Ac{\'\i}n}},
  \bibinfo{journal}{Physical review letters} \textbf{\bibinfo{volume}{111}},
  \bibinfo{pages}{240401} (\bibinfo{year}{2013}).

\bibitem[{\citenamefont{Giorgi and Campbell}(2015)}]{giorgi2015correlation}
\bibinfo{author}{\bibfnamefont{G.~L.} \bibnamefont{Giorgi}} \bibnamefont{and}
  \bibinfo{author}{\bibfnamefont{S.}~\bibnamefont{Campbell}},
  \bibinfo{journal}{Journal of Physics B: Atomic, Molecular and Optical
  Physics} \textbf{\bibinfo{volume}{48}}, \bibinfo{pages}{035501}
  (\bibinfo{year}{2015}).

\bibitem[{\citenamefont{Spohn}(1978)}]{spohn1978b}
\bibinfo{author}{\bibfnamefont{H.}~\bibnamefont{Spohn}},
  \bibinfo{journal}{Journal of Mathematical Phyics}
  \textbf{\bibinfo{volume}{19}}, \bibinfo{pages}{1227} (\bibinfo{year}{1978}).

\bibitem[{\citenamefont{Esposito et~al.}(2010)\citenamefont{Esposito,
  Lindenberg, and den Broeck}}]{esposito2010b}
\bibinfo{author}{\bibfnamefont{M.}~\bibnamefont{Esposito}},
  \bibinfo{author}{\bibfnamefont{K.}~\bibnamefont{Lindenberg}},
  \bibnamefont{and} \bibinfo{author}{\bibfnamefont{C.~V.} \bibnamefont{den
  Broeck}}, \bibinfo{journal}{New Journal of Physics}
  \textbf{\bibinfo{volume}{12}}, \bibinfo{pages}{013013}
  (\bibinfo{year}{2010}).

\end{thebibliography}

\end{document}